\documentclass[aps,prl,twocolumn]{revtex4}
\usepackage{graphicx}
\usepackage{latexsym}
\usepackage{amsmath}
\usepackage{amsfonts}
\usepackage{amssymb}
\usepackage{color}
\usepackage{units}
\usepackage{natbib}
\newcommand{\be}{\begin{equation}}
\newcommand{\ee}{\end{equation}}
\newcommand{\bea}{\begin{eqnarray}}
\newcommand{\eea}{\end{eqnarray}}
\usepackage[parfill]{parskip}
\usepackage{orcidlink}
\begin{document}
\title{Multiple pattern formation in quorum sensing of density-enhanced motility} 
\author{Itay Azizi \orcidlink{0000-0003-2939-4421}}
\affiliation{Independent Researcher, Vitoria-Gasteiz, Spain}
\email{itay.azizi@gmail.com} 
\begin{abstract} 
Using Langevin dynamics simulations, I investigate nonequilibrium systems of particles following a density-enhanced motility (DEM) rule: particles are passive below a critical local density and active above it. This mechanism represents an inverse of the conventional quorum-sensing rule. I explore specifically quorums much larger than particle size and at two levels of activity. Above critical values of the density, quorum size, and activity, the system undergoes phase separation into low- and high-energy regions. The passive particles organize into distinct spatial patterns, including holes, stripes, and labyrinthine structures, while the active particles form a gas. I characterize the resulting steady states and identify the qualitative mechanisms governing the selection of different morphological regimes. I further discuss the relevance of these results to biological systems governed by analogous quorum-sensing mechanisms and to biological systems exhibiting related forms of spatial organization. Finally, I present directions for future investigation.
\end{abstract} 
\maketitle 
\section{Introduction}
Within the framework of intelligent active matter, growing interest in the soft-matter community has focused on materials that respond or adapt to their local environment. One class of models captures such environmental responsiveness through a coupling between particle motility and local density. This coupling has been studied extensively for the case in which activity decreases with increasing density \cite{Bauerle2018,Fischer2020,Jose2021} and, more recently, for the inverse case in which activity increases with increasing density \cite{Azizi2026}. In the conventional case, particles become passive in dense regions and remain active in dilute regions. At sufficiently high activity, this mechanism can drive phase separation into a dense phase of passive particles coexisting with a dilute fluid of active particles. 
Depending on the model parameters, the dense phase may be fluid, amorphous, or crystalline, and may exhibit a variety of cluster morphologies \cite{Azizi2026,Souza2025}.
\\In biological systems, quorum sensing is considerably more complex than the simplified mechanisms typically considered in physics models. Importantly, the spatial organization generated by quorum sensing can have direct functional consequences. A densely packed or solid-like group may suppress the exchange of materials and information between individuals, whereas a fluid-like configuration can facilitate such exchange and support collective and cooperative behavior. The morphology and mechanical state of an aggregate can therefore influence its functionality and survival.
\\Several biological systems exhibit forms of density-enhanced motility. Locusts, for example, undergo a density-dependent behavioral transition \cite{Buhl2006}: isolated individuals exhibit a solitarious phase characterized by relatively low activity, whereas sufficiently high local density induces a gregarious phase with enhanced activity and collective alignment. This transition is mediated in part by tactile stimulation between conspecifics, which increases serotonin levels and promotes the behavioral switch. Myxobacteria and other social bacteria likewise exhibit density-dependent increases in motility mediated by quorum sensing \cite{Daniels2004}; the accumulation of chemical signals above a threshold can upregulate motility-related genes and promote collective swarming. In certain species, sperm cells also exhibit enhanced motility at high local densities through hydrodynamic coupling: aggregation into trains or bundles can reduce effective drag through collective flow fields and thereby increase swimming speed \cite{Moore2002}. Army ants provide another example, in which local crowding, together with pheromonal and tactile cues, can induce a transition from relatively slow exploratory motion to highly motile and aligned trail-following behavior \cite{Couzin2003}. 
\\Human crowd dynamics under panic conditions have also been modeled as exhibiting a regime in which increasing density is accompanied by increased velocity and forceful pushing, in contrast to the jamming typically associated with high density. This regime, however, remains strongly context-dependent and experimentally contested \cite{Helbing2000}. A natural question is therefore how the characteristic length scale of the quorum influences the collective states generated by density-enhanced motility. This is motivated by the fact that the interactions underlying quorum sensing—such as visual perception—can be long-ranged, suggesting a link between the range of "sight" and the resulting global organization. I investigate this question across a broad range of quorum sizes and global densities, and for two levels of activity. A threshold density is defined that determines whether a particle responds to the local density relative to the ambient mean density. To establish a well-defined microscopic length scale, particles are modeled as interacting through an approximate hard-core potential.
\\Accordingly, I ask: how does quorum length scale affect the emergent states of a system of hard particles whose motility is enhanced by local density?
\section{Methods}
\begin{figure}[ht] 
\includegraphics[width=0.95\linewidth]{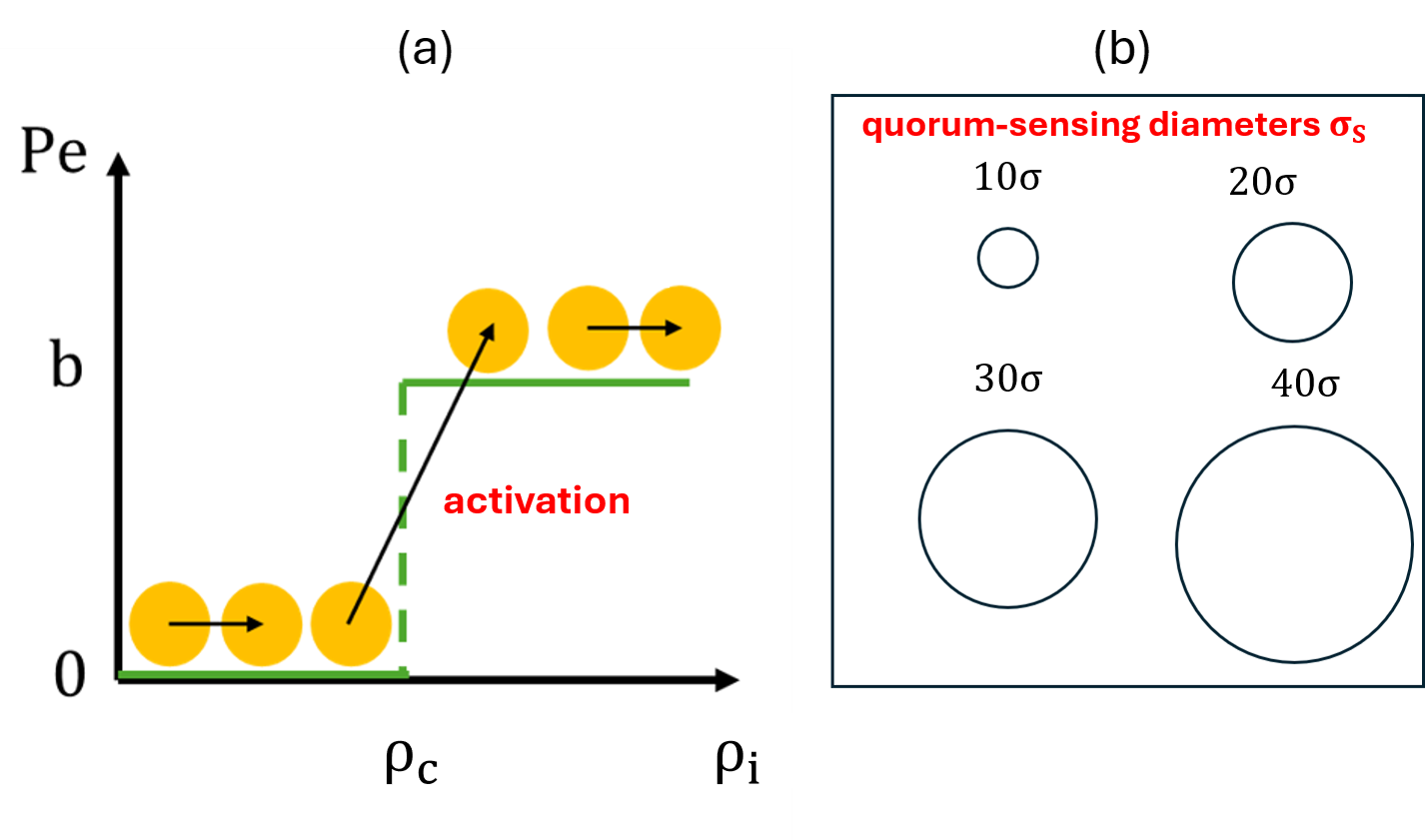}
	\caption{\label{fig:patterns} (a) Density-enhanced motility rule: when the local density $\rho_i$ of particle $i$ crosses a critical values $\rho_c$, the particle is activated. (b) Different quorum sizes: $\sigma_{s}$. $\rho_i$ is always measured over a disk with $\sigma_{S}$ diameter.}
\end{figure}
In order to elucidate the behavior of systems exhibiting density-enhanced  motility, Langevin dynamics simulations were carried out using a custom code written in Fortran. The two-dimensional system consists of $N$ particles in a square box with periodic boundary conditions and dimensions $L=L_x=L_y=100\sigma$ which sets a global density of $\rho=N/L^2$. 
\\Particles $i$ and $j$ interact via a Weeks-Chandler-Andersen short range repulsive potential 
\begin{equation}
V(r_{ij})=4\epsilon[(\sigma/r_{ij})^{12}-(\sigma/r_{ij})^{6}]+\epsilon
\end{equation}
where cutoff distance is $r_c=2^{1/6}\sigma$, $\mathbf{r}_{ij}=\lvert\mathbf{r}_j-\mathbf{r}_i\rvert$ and $\epsilon=1$.
\\The choice of a purely repulsive potential ensures that any observed clustering behavior can be attributed to quorum sensing rather than to direct attractive interactions. I set $k_B=1$ and $T=0.02$; such a low temperature yields approximately hard-core interactions, with $\sigma$ the unit of length, approximately equal to the particle diameter.
Particle $i$ alternates between active and passive identity as a function of its local density $\rho_i$. The code computes $m_i$, the number of neighbors of particle $i$ within a disk of diameter $\sigma_s$ centered on particle $i$, where $\sigma_s = 2R_s$; thus the local density is $\rho_i = m_i/\pi R_s^2$. The global and critical densities are taken to be equal, so that the particle responds to deviations from the ambient density: $\rho_g = \rho_c = \rho$.
\\Particle $i$ is assigned a position $r_i$ and an orientation $\theta_i$, which evolve according to the following Langevin equations:
\begin{eqnarray}
\frac{d\mathbf{r}_i(t)}{dt} &=& v_p(\rho_i) \mathbf{e_i}(t) - \beta D \nabla U_i(t) + \sqrt{2D} \, \boldsymbol{\eta}(t), \label{eom1} \\
\frac{d\theta_i(t)}{dt} &=& \sqrt{2D_r(\rho_i)} \, \xi(t)
\label{eom2}
\end{eqnarray}
where $v_p$ is the magnitude of the self-propulsion velocity, oriented along $\mathbf{e}_i(t) = (\cos\theta_i(t), \sin\theta_i(t))$. The Péclet number ($\text{Pe}$) is defined as the ratio of the persistence length $l_p = v_p \tau_r$ — the distance an active particle travels before its orientation randomizes due to rotational diffusion, with $\tau_r = 1/D_r$ — to the diffusive length $l_D = \sqrt{D \tau_r}$ — the distance the particle diffuses over this same time interval.
\begin{equation}
\text{Pe} = \dfrac{l_p}{l_D} = \dfrac{v_p}{\sqrt{DD_r}}.
\label{pe}
\end{equation}
\\The translational (rotational) diffusion coefficient is denoted by $D$ ($D_r$), and $\boldsymbol{\eta}(t)$ and $\xi(t)$ are Gaussian white noise terms satisfying $\langle \eta_i(t) \eta_j(t^{\prime}) \rangle = \delta_{ij}\delta(t-t^{\prime})$, with $i,j \in (x,y)$, and $\langle \xi(t)\xi(t^{\prime}) \rangle = \delta(t-t^{\prime})$, respectively. $\gamma$ and $\gamma_r$ denot{e the translational and rotational friction coefficients, satisfying $D = k_BT/\gamma$ and $D_r = k_BT/\gamma_r$, respectively, with $\gamma = 2$ and $\gamma_r = 0.01$.
$U$ denotes the potential energy arising from conservative forces, e.g., pair interactions such that $U_i = \sum_{j \neq i} V(r_{ij})$, where the gradient is taken with respect to the position of particle $i$. 
According to local density, a passive particle is assigned $v_p = \text{Pe} = D_r = 0$, while an active particle is assigned $v_p$ such that $\text{Pe} = b$.  
The constants for all systems studied are $L, T, \gamma, \gamma_r$, denoting the box length, temperature, translational friction coefficient, and rotational friction coefficient, respectively. The variables are $\rho, R_s$, denoting the density and sensing radius, respectively. Thus, each simulation studies a single point in this $(\rho, R_S)$ phase space.
\\Each simulation begins from a square lattice and is run for up to $10^5$ steps with a timestep of $\Delta t = 0.005$, which is sufficient for relaxation, as monitored by plateaus in the potential energy and active particle fraction, and by small variation in cluster size and morphology. Having worked at a single activity of $b=10$, I additionally examined a higher value, $b=40$.
\section{Results}
Since particles continuously switch between passive and active states in response to their local density, the microscopic configuration of the system is inherently heterogeneous and time-dependent. Rather than characterizing the system through conventional static or dynamical observables defined for particles of fixed identity, I therefore focus on the emergent global morphology and spatial organization that result from this switching behavior. Specifically, I characterize the distinct steady-state morphologies arising as a function of quorum size, global density, and activity. The Results section is organized as follows: Section~1 presents a visual characterization of the steady states; Section~2 establishes the relation between the emergent morphology and the control parameters; and Section~3 examines the dependence of spatial ordering on global density.
\subsubsection{Visual inspection}
I first examine the steady-state configurations obtained for different quorum sizes and activity levels. The simulations reveal several distinct morphological regimes, including holes, stripes, and labyrinthine patterns (Figs.~\ref{fig:b10} and \ref{fig:b40}). An important point concerning the interpretation of these configurations is that visually dense regions are not necessarily active. Because the sensing diameter can substantially exceed the characteristic interparticle spacing, the local density relevant for determining motility differs from the instantaneous particle coordination within a visually dense region. Thus, regions that appear dense in the snapshots can remain below the sensing threshold and therefore consist predominantly of passive particles.
\begin{figure}[ht]
\centering
\includegraphics[width=0.95\linewidth]{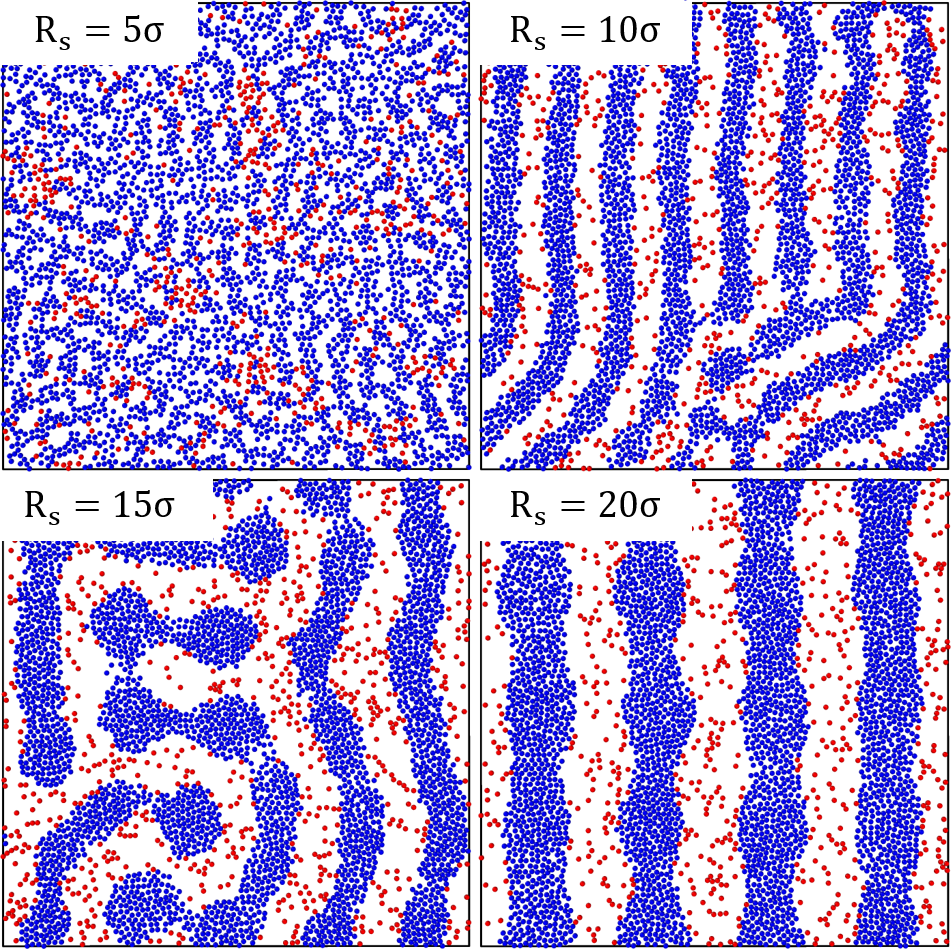}
\caption{Steady states at global density $\rho=0.4$, constant activity $b=10$, and varying quorum size $R_{s}$. Passive (active) particles are shown in blue (red). As the quorum size increases, passive particles aggregate into increasingly extended domains and organize into stripe-like structures coexisting with a gas of active particles.}
\label{fig:b10}
\end{figure}
\\At $R_{s}=5\sigma$, corresponding to a short sensing range, passive and active particles remain largely intermixed, with only weak and small-scale clustering. No pronounced large-scale organization is observed, consistent with the limited spatial range over which particles can respond to their surroundings.
At $R_{s}=10\sigma$, distinct phase separation begins to emerge. Passive particles condense into elongated, wavy bands and filaments, while active particles accumulate around and between these domains. The resulting stripes remain relatively thin, curved, and irregular, with multiple branches.
At $R_{s}=15\sigma$, the passive phase undergoes further coarsening. The stripes thicken and locally merge into larger, more compact domains, while the characteristic domain size increases relative to the $R_{s}=10\sigma$ case. At $R_{s}=20\sigma$, the system reaches its most strongly coarsened morphology. Passive particles form thick, relatively straight, well-separated bands that span the simulation box, while active particles accumulate predominantly along the interfaces. Visual inspection indicates that the interfacial roughness decreases with increasing $R_{s}$.
\\Overall, increasing $R_{s}$ extends the spatial range over which local density influences particle motility, thereby promoting larger-scale phase separation and domain coarsening. The characteristic pattern length consequently increases with the sensing range, producing a progression from weakly organized, fine-scale structures to well-defined, macroscopically separated bands. 
\\Simulations initialized from different initial configurations, including phase separated case, converged to qualitatively similar steady states, indicating that the observed morphologies are robust and not strongly dependent on the initial condition. This robustness suggests that the identified morphologies represent genuine outcomes of the underlying dynamics rather than artifacts of a particular preparation protocol. Moreover, the convergence to consistent steady states across independent realizations, despite the stochastic nature of the active dynamics and the intrinsic heterogeneity introduced by the switching between passive and active identities, supports the interpretation of these morphologies as generic outcomes of the density-dependent motility mechanism, rather than being contingent on fine-tuned or transient microscopic arrangements.
\begin{figure}[ht]
\centering
\includegraphics[width=0.95\linewidth]{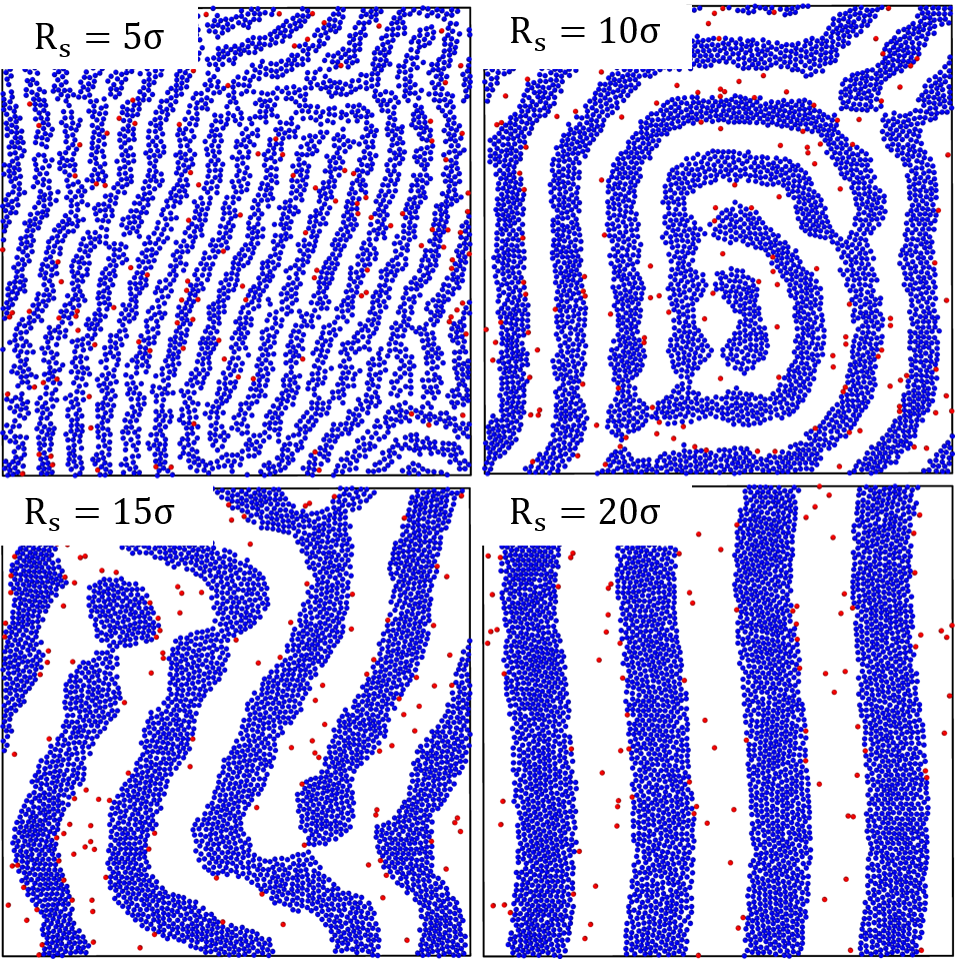}
\caption{Steady states at global density $\rho=0.4$, constant activity $b=40$, and varying quorum size $R_{s}$. Passive (active) particles are shown in blue (red). Increasing the quorum size promotes the formation and coarsening of passive domains, which organize the passive particles into labyrinthine structures and, at larger quorum sizes, into stripe-like structures, coexisting with a gas of active particles.}
\label{fig:b40}
\end{figure}
\\At the higher activity, $b=40$, shown in Fig.~\ref{fig:b40}, spatial organization again develops progressively with increasing $R_{S}$. Relative to the lower-activity case, the passive domains are more compact and their interfaces are visibly smoother. The degree of local hexatic order is also enhanced, suggesting that increased activity promotes stronger ordering within the passive domains.
\\Considered jointly, these observations indicate that both the sensing range and the activity jointly control the emergent morphology. A finite activity is required for robust pattern formation: below a threshold activity, the system remains comparatively homogeneous (not shown). This threshold is attributed to the requirement that particles persist in the active state sufficiently long to establish and maintain stable quorums of adequate size. When the persistence length is too short, density fluctuations are rapidly disrupted, favoring a well-mixed state. The number of active particles further influences the relaxation dynamics: when active particles are abundant, the passive domains retain a degree of internal motion, whereas when active particles are scarce, the passive stripes form a stable solid.
\subsubsection{Relation between the emergent morphology and the control parameters}
For $\rho=0.4$, the simulations reveal a systematic correspondence between the initial control parameters and the resulting morphology. Hole-like patterns occur predominantly at low sensing radius and low activity, whereas stripe-like patterns emerge when both quantities are sufficiently large. Labyrinthine morphologies are favored by the combination of a relatively short sensing range and high activity. The change from holes to stripes happens as activity increases and then the active matter can break the "bridges" of the hole pattern and create stripes. To conclude, using these qualitative reasoning, the particular morphology can be roughly anticipated from the location of the system in the $(R_{s},b)$ parameter space. Of course, at a very high activity, numerical problems will arise and the hard-core approximation will break down (not shown).
\subsubsection{Dependence of spatial ordering on global density}
I next examine the effect of global density at fixed activity $b=10$ and sensing range $R_s=20\sigma$ at Figure~\ref{fig:wide} which shows snapshots of only the passive particles for densities ranging from $\rho=0.3$ to $0.6$, revealing a systematic and striking increase in spatial ordering with increasing density. Passive particles are colored according to their local hexatic order parameter $\psi_6$, and the average value $\langle\psi_6\rangle$ over all passive particles is indicated above each snapshot and increases monotonically with $\rho$. This growing structural order is accompanied by a marked sharpening of the interfaces separating passive and active regions, which become progressively flatter and smoother as the global density increases.
\begin{figure*}[t]
\centering
\includegraphics[width=1.0\textwidth]{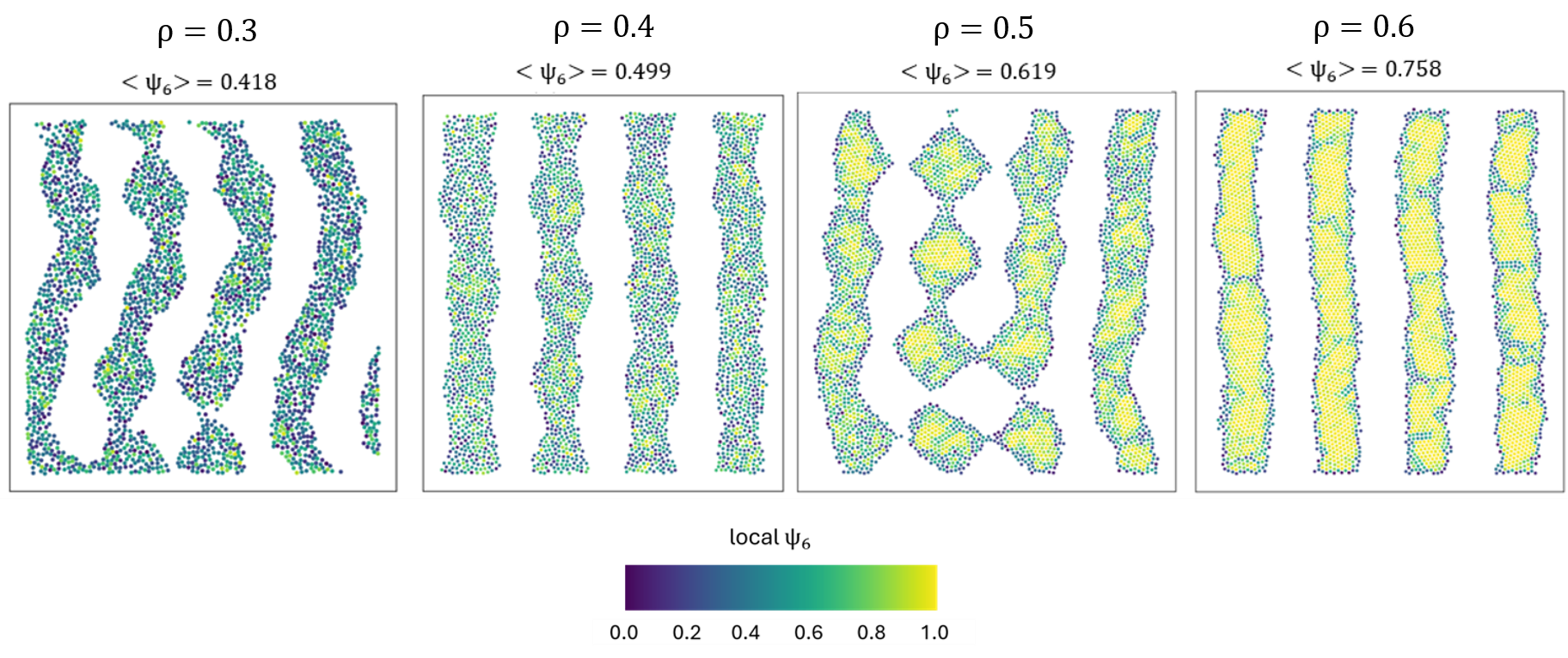}
\caption{Steady states of only passive particles at activity $b=10$ and quorum size $R_s=20\sigma$ for increasing global density in the range $\rho=0.3$--$0.6$. Particle color indicates local hexatic order, according to the colormap shown below the snapshots. Increasing global density produces more compact passive domains with smoother interfaces, accompanied by enhanced local hexatic ordering.}
\label{fig:wide}
\end{figure*}
\\These results establish global density as an additional control parameter governing the morphology and internal organization of the passive phase: increasing density drives more compact domains, reduced interfacial roughness, and enhanced structural order. 
\\Together with the sensing range and activity, density thus completes a three-parameter picture in which emergent morphology is governed jointly by the quorum length scale, motility, and global density.
\section{Discussion}
I have studied a minimal model of density-enhanced motility, in which activity increases, rather than decreases, with local density. I find that this reversal of the conventional density--motility relation is sufficient to generate critical behavior and spontaneous spatial organization. To the best of my knowledge, only one prior theoretical study of density-enhanced motility has reported pattern formation, via an analytical treatment \cite{Ridgway2023}. 
\\The present results extend this direction by demonstrating that, for finite sensing ranges, density-enhanced activation gives rise to a rich variety of distinct morphologies, including holes, stripes, and labyrinthine structures.
\\The mechanism is based on a positive feedback between local density and activity, a coupling that is notable for acting over a long range rather than through direct, short-range interactions. This feedback is qualitatively different from the negative density--motility coupling underlying the more commonly studied forms of motility-induced phase separation. 
\\The present results suggest that the mathematical structure of density-dependent feedback may be relevant beyond biological active matter. Systems as different as ecological populations, engineered active agents, and human crowds can exhibit collective behavior in which local population density modifies the activity or response of individuals. In human systems, for example, density can influence behavior through visual, social, and communicative interactions operating over length scales much larger than the individual. Such analogies should not be interpreted as evidence for a common microscopic mechanism. Rather, they indicate that quorum-like feedback may provide a useful abstract framework for comparing collective organization across physical, biological, and social systems.
\\ Pattern formation driven by density-dependent feedbacks has also been widely documented in ecological systems. In particular, theoretical and experimental studies of vegetation dynamics have shown that local interactions between population density, resource availability, and transport processes can generate a variety of spatial structures, including spots, stripes, and labyrinthine patterns \cite{Clerc2021,Meron2004}. These ecological examples demonstrate that relatively simple density-dependent feedback mechanisms can lead to spontaneous spatial organization and pattern selection. Although the microscopic mechanisms in our model are different, the emergence of holes, stripes, and labyrinthine structures from a density-dependent motility feedback places our results within this broader context of self-organized pattern formation in biological and ecological systems.
\\The model also leaves several physical questions open. The present study is restricted to two-dimensional particles interacting through an approximate hard-core potential. The observed patterns may therefore depend on the interaction potential and dimensionality. In particular, simulations with softer interactions did not produce comparably stable patterns \cite{Azizi2026}, suggesting that steric interactions may play an important role in stabilizing the morphologies reported here. A systematic comparison of interaction potentials would therefore be useful for separating generic features of density-enhanced activation from effects associated with the microscopic interaction law.
\\Further extensions could consider three-dimensional systems, anisotropic or deformable particles, and mixtures of particles with different sizes or sensing properties. Particle shape may introduce additional orientational degrees of freedom and generate structures not accessible in the present model. Indeed, quorum-sensing models involving rod-like particles have been shown to produce structures such as asters and stripes \cite{Velasco2018}, while size-disparate mixtures can introduce additional effective interactions through depletion mechanisms \cite{Azizi2025}. These extensions could substantially enlarge the range of possible steady states.
\\More broadly, the results indicate that quorum sensing can be viewed not only as a biological communication mechanism but also as a general route to nonequilibrium self-organization. The present model is deliberately minimal, and its purpose is to identify the consequences of reversing the conventional density dependence of motility. Its main contribution is therefore not a specific prediction for a particular biological system, but the identification of a simple physical mechanism capable of producing criticality and multiple spatial morphologies. Developing richer models in which physical motion, chemical signaling, biological regulation, and, where appropriate, social interactions are treated within a common framework could provide a systematic route toward understanding how local sensing rules generate collective organization across different classes of systems.
\section{Acknowledgments}
The author acknowledges useful discussions with Ehud Meron, Takeaki Araki and Cesare Nardini. This work was self-funded by the author.
\bibliography{references}
%\printbibliography
\end{document}